\documentclass[11pt]{article}
\usepackage{amsmath,amssymb}
\usepackage[dvips]{epsfig}
\newtheorem{theorem}{Theorem}[section]

\newtheorem{corollary}[theorem]{Corollary}

\newtheorem{definition}{Definition}[section]
\newtheorem{example}[theorem]{Example}
\newtheorem{remark}[theorem]{Remark}
\protect \hoffset -1 true cm \voffset -1.5 true cm \textwidth 15cm
\newcommand{\vu}{\boldsymbol{u}}
\newcommand{\ru}{\boldsymbol{r}}
\newcommand{\no}{\boldsymbol{n}}

\begin{document}

\title{Reciprocal transformations and local Hamiltonian structures
of hydrodynamic type systems}

\author {Simonetta Abenda \\
Dipartimento di Matematica and C.I.R.A.M. \\
Universit\`a degli Studi di Bologna, Italy \\
{\footnotesize  abenda@ciram.unibo.it} } \maketitle
\begin{abstract}
We start from a hyperbolic DN hydrodynamic type system of
dimension $n$ which possesses Riemann invariants and we settle the
necessary conditions on the conservation laws in the reciprocal
transformation so that, after such a transformation of the
independent variables, one of the metrics associated to the
initial system be flat. We prove the following statement: let
$n\ge 3$ in the case of reciprocal transformations of a single
independent variable or $n\ge 5$ in the case of transformations of
both the independent variable; then the reciprocal metric may be
flat only if the conservation laws in the transformation are
linear combinations of the canonical densities of conservation
laws, {\it i.e} the Casimirs, the momentum and the Hamiltonian
densities associated to the Hamiltonian operator for the initial
metric. Then, we restrict ourselves to the case in which the
initial metric is either flat or of constant curvature and we
classify the reciprocal transformations of one or both the
independent variables so that the reciprocal metric is flat. Such
characterization has an interesting geometric interpretation: the
hypersurfaces of two diagonalizable DN systems of dimension $n\ge
5$ are Lie equivalent if and only if the corresponding local
hamiltonian structures are related by a canonical reciprocal
transformation.
\end{abstract}

\section{Introduction}

Systems of hydrodynamic type are quasilinear evolutionary
hyperbolic PDEs of the form
\begin{equation}\label{abe:DN}
u^i_t = \sum_{k=1}^n v^i_k (u) u^k_x,\quad\quad
u=(u^1,\dots,u^n),\quad u^i_x =\frac{\partial u^i}{\partial
x},\quad u^i_t =\frac{\partial u^i}{\partial t}.
\end{equation}
They naturally arise in applications such as gas dynamics,
hydrodynamics, chemical kinetics, the Whitham averaging procedure,
differential geometry and topological field
theory~\cite{abe:DN83,abe:DN89,abe:D96,abe:T85,abe:T91}. Dubrovin
and Novikov~\cite{abe:DN83} showed that (\ref{abe:DN}) is a local
Hamiltonian system (DN system) with Hamiltonian $H[u] = \int h(u)
dx$, if there exists a flat non-degenerate metric tensor $g(u)$ in
${\mathbb R}^n$ with Christoffel symbols $\Gamma^i_{jk} (u)$, such
that the matrix $v^i_k(u)$ can be represented in the form
\begin{equation}\label{abe:ham1}
v^i_k (u)= \sum_{k=1}^n \left( g^{il} (u) \frac{\partial^2
h}{\partial u^l\partial u^k}(u) -\sum_{s=1}^n g^{ik}(u)
\Gamma^l_{sk} (u)\frac{\partial h}{\partial u^l}(u)\right).
\end{equation}
In this paper we shall consider DN systems which possess Riemann
invariants, {\it i.e.} they may be transformed to the diagonal
form
\begin{equation} u^i_t = v^i(\vu) u^i_x,\quad\quad
i=1,\dots,n,
\end{equation}
with $n\ge 3$ and with $v^i(\vu)$ all real and distinct (strict
hyperbolicity property). We also suppose to work in the space of
smooth and rapidly decreasing functions so that
$(\frac{d}{dx})^{-1} f_x =f$.

If $n=2$, (\ref{abe:DN}) can always be put in diagonal form and
are integrable by the hodograph method. For arbitrary $n$,
Tsarev~\cite{abe:T85} proved that a DN system as in
(\ref{abe:DN}), (\ref{abe:ham1}) can be integrated by a
generalized hodograph method only if it may be transformed to the
diagonal form. In the latter case, moreover the flat metric is
diagonal, the Hamiltonian satisfies
\begin{equation}\label{abe:conslaw}
\frac{\partial^2 h}{\partial u^i\partial u^j} = \Gamma^{i}_{ij}
(u) \frac{\partial h}{\partial u^i} + \Gamma^{j}_{ji} (u)
\frac{\partial h}{\partial u^j},
\end{equation}
and each solution to (\ref{abe:conslaw}) generates a conserved
quantity for the DN system (\ref{abe:DN}), (\ref{abe:ham1}) and
all Hamiltonian flows generated by these conserved densities
pairwise commute. As a consequence, for $n\ge 3$, DN systems which
possess Riemann invariants are always integrable. We recall that
there do also exist DN systems with an infinite number of
conserved quantities which do not possess Riemann invariants (see
Ferapontov \cite{abe:F95-1} for the classification of the latter
when $n=3$).

Since a non-degenerate flat diagonal metric in ${\mathbb R}^n$ is
associated to an orthogonal coordinate system $u^i = u^i
(x^1,\dots,x^n)$, there is a natural link between diagonalizable
Hamiltonian systems and $n$--orthogonal curvilinear coordinates in
flat spaces. Upon introducing the Lam\'e coefficients, which in
our case take the form
\[
\displaystyle H_i^2 (u) = \sum_{k} \left( \frac{\partial
x^i}{\partial u^k} \right)^2,
\]
the metric tensor in the coordinate system $u^i$ is diagonal $ds^2
= \sum\limits_{i=1}^n H_i^2 (u) (du^i)^2$, and the zero curvature
conditions $R_{il,im}(u) =0$ ($i\not = l \not = m \not = i$) and
$R_{il,il}(u)=0$ ($i\not = l$) form an overdetermined system:
\begin{eqnarray}
\frac{\partial^2 H_i}{\partial u^l\partial u^m} &=& \frac{1}{H_l}
\frac{\partial H_l}{\partial u^m}\frac{\partial H_i}{\partial u^l}
+\frac{1}{H_m} \frac{\partial H_m}{\partial u^l}\frac{\partial
H_i}{\partial u^m},\label{abe:darboux1}
\\
\frac{\partial}{\partial u^l} \frac{\partial H_i}{H_l\partial
u^l}&+&\frac{\partial}{\partial u^i}\frac{\partial
H_l}{H_i\partial u^i}+\sum_{m\not = i,l}
\frac{1}{H_m^2}\frac{\partial H_i}{\partial u^m}\frac{\partial
H_l}{\partial u^m}=0.\label{abe:darboux2}
\end{eqnarray}
Bianchi and Cartan showed that a general solution to the zero
curvature equations (\ref{abe:darboux1}), (\ref{abe:darboux2}) can
be parametrized locally by $n(n-1)/2$ arbitrary functions of two
variables. If the Lam\'e coefficients $H_i(u)$ are known, one can
find $x^i(u^1,\dots,u^n)$ solving the linear overdetermined
problem (embedding equations)
\begin{equation}\label{abe:emb}
\frac{\partial^2 x^i}{\partial u^k\partial u^l} = \Gamma^{k}_{kl}
(u) \frac{\partial x^i}{\partial u^k} + \Gamma^{l}_{lk} (u)
\frac{\partial x^i}{\partial u^l}, \quad\quad \frac{\partial^2
x^i}{\partial (u^l)^2} =\sum_k \Gamma^{k}_{ll} (u) \frac{\partial
x^i}{\partial u^k}.
\end{equation}
Comparison of Eqs. (\ref{abe:conslaw}) and (\ref{abe:emb}) implies
that the flat coordinates for the metric $g_{ii} (u) = (H^i(u))^2$
are the Casimirs of the corresponding Hamiltonian operator.
Finally, Zakharov~\cite{abe:Z98} showed that the dressing method
may be used to determine the solutions to the zero curvature
equations up to Combescure transformations.

It then follows that the classification of flat diagonal metrics $
ds^2 = g_{ii}( u) (du^i)^2 $ is an important preliminary step in
the classification of integrable Hamiltonian systems of
hydrodynamic type. Best known examples of integrable Hamiltonian
systems of hydrodynamic type possess Riemann invariants, a pair of
compatible flat metrics and have been obtained in the framework of
semisimple Frobenius manifolds (axiomatic theory of integrable
Hamiltonian systems) \cite{abe:D96,abe:D98H,abe:D98}; in the
latter case, one of the flat metrics is also Egorov ({\it i.e.}
its rotation coefficients are symmetric).

\medskip

Reciprocal transformations change the independent variables of a
system and are an important class of nonlocal transformations
which act on hydrodynamic--type
systems~\cite{abe:RS82,abe:P95,abe:F95,abe:FP03,abe:AG05,
abe:XZ06,abe:AG07}. Reciprocal transformations map conservation
laws to conservation laws and map diagonalizable systems to
diagonalizable systems, but act non trivially on the metrics and
on the  Hamiltonian structures: for instance, the flatness
property and the Egorov property for metrics as well as the
locality of the Hamiltonian structure  are not preserved, in
general, by such transformations. Then, it is natural to
investigate under which additional hypotheses the reciprocal
system still possesses a local Hamiltonian structure, our ultimate
goal being the search for new examples of integrable Hamiltonian
systems and the geometrical characterization of the associated
hypersurfaces.

With this in mind, in the following we start from a smooth
integrable Hamiltonian system in Riemann invariant form
\begin{equation}\label{abe:DNriem}
u^i_t= v^i(u)u^i_x,\quad\quad i=1,\dots,n,
\end{equation}
with smooth conservation laws
\begin{equation}\label{abe:consrec}
B(u)_t=A(u)_x,\quad N(u)_t=M(u)_x \end{equation} with
$B(u)M(u)-A(u)N(u)\neq 0$. In the new independent variables ${\hat
x}$ and ${\hat t}$ defined by
\begin{equation}\label{abe:grarec}
d{\hat x} = B (u)dx + A(u)dt,\quad\quad d{\hat t} = N(u)dx
+M(u)dt,
\end{equation}
the reciprocal system is still diagonal and takes the form
\begin{equation}\label{abe:rec}
u^{i}_{{\hat t}}
=\frac{B(u)v^i(u)-A(u)}{M(u)-N(u)v^i(u)}u^{i}_{{\hat x}}={\hat
v}^i (u) u^{i}_{{\hat x}}.
\end{equation}
Moreover, the metric of the initial systems $g_{ii}(u)$ transforms
to
\begin{equation}\label{abe:recmetint}
{\hat g}_{ii}(u)= \left(\frac{M(u)-N(u)v^i(u)}{B(u)M(u)-A(u)N(u)}
\right)^2{ g}_{ii}(u),
\end{equation}
and all conservation laws and commuting flows of the original
system (\ref{abe:DNriem}) may be recalculated  in the new
independent variables.

If the reciprocal transformation is linear ({\it i.e.} ${A,B,N,M}$
are constant functions), then the reciprocal to a flat metric is
still flat and locality and compatibility of the associated
Hamiltonian structures are preserved (see
Refs.~\cite{abe:T91,abe:P95,abe:XZ06}).

Under a general reciprocal transformation, the Hamiltonian
structure does not behave trivially and a thorough study of
reciprocal Hamiltonian structures is still an open problem.
Ferapontov and Pavlov~\cite{abe:FP03} construct the reciprocal
Riemannian curvature tensor and the reciprocal Hamiltonian
operator when the initial metric is flat, while in
\cite{abe:AG07}, we construct the reciprocal Riemannian curvature
tensor and the reciprocal Hamiltonian operator when the initial
metric is not flat and the initial system also possesses a flat
metric.

The classification of the reciprocal Hamiltonian structures is
also complicated by the fact that a DN system as
in~(\ref{abe:DN})-(\ref{abe:ham1}) also possesses an infinite
number of nonlocal Hamiltonian
structures~\cite{abe:FM90,abe:F95,abe:MN01,abe:Ma05}. It is then
possible that two DN systems are linked by a reciprocal
transformation and that the flat metrics of the first system are
not reciprocal to the flat metrics of the second. In
\cite{abe:AG05}, we constructed such an example: the genus one
modulation (Whitham-CH) equations associated to Camassa-Holm in
Riemann invariant form ($n=3$ in (\ref{abe:DNriem})). We proved
that the Whitham-CH equations are a DN-system and possess a pair
of compatible flat metrics (none of the metrics is Egorov). We
also proved the connection via a reciprocal transformation of the
Whitham-CH equations to the modulation equations associated to the
first negative flow of the Korteweg de Vries hierarchy
(Whitham-KdV$_{-1}$). In \cite{abe:AG05}, finally we also found
the relation between the Poisson structures of the
Whitham-KdV$_{-1}$ and the Whitham-CH equations: both systems
possess a pair of compatible flat metrics, and the two flat
metrics of the first system are respectively reciprocal to the
constant curvature and conformally flat metrics of the second (and
vice versa).

In view of the above results, in~\cite{abe:AG07} we have started
to classify the reciprocal transformations which transform a DN
system to a DN system, under the condition that the flat metric
tensor ${\hat g}(u)$ of the transformed system is reciprocal to a
metric tensor $g(u)$ of the initial system, which is either flat
or of constant Riemannian curvature or conformally flat.

In~\cite{abe:AG07}, we give necessary and sufficient conditions so
that a reciprocal transformation which changes only one
independent variable may preserve the flatness of the metric; in
particular, we show that the conservation laws in the reciprocal
transformation of the independent variable $x$ (resp. $t$) are
linear combinations of Casimirs and momentum densities (resp.
Casimirs and Hamiltonian densities).

For an easier comparison with the results known in literature, we
recall that Ferapontov~\cite{abe:F95} takes a reciprocal
transformation where the conservation laws in (\ref{abe:grarec})
are a linear combination of the Casimirs, momentum and Hamiltonian
densities and gives the necessary and sufficient conditions so
that starting from a flat metric $g (u)$, the reciprocal metric
${\hat g}(\vu)$ be either a flat or a constant curvature metric.
Following Ferapontov \cite{abe:F95-1,abe:F95}, we call canonical a
reciprocal transformation in which the integrals in
(\ref{abe:grarec}) are linear combinations of the $n+2$ canonical
integrals (Casimirs, Hamiltonian and momentum) with respect to the
given Hamiltonian structure.

The results in ~\cite{abe:AG07,abe:F95} suggest that canonical
reciprocal transformations have a privileged role in preserving
locality of the Hamiltonian structure. In this paper we show that
canonical transformations are indeed the only reciprocal
transformations which may transform the initial metric $g_{ii}
(\vu)$ into a reciprocal flat metric ${\hat g}_{ii} (\vu)$ when
the dimension of the system is $n\ge 3$ (in the case of a
transformation of a single independent variable) or $n\ge 5$ (in
the case of a transformation of both the independent variables).

First of all, in Theorems \ref{abe:teonec1} and \ref{abe:teonec2},
we give necessary conditions on the initial metric $g_{ii}(\vu)$
and on the conservation laws (\ref{abe:consrec}) in the reciprocal
transformation, so that the reciprocal metric
(\ref{abe:recmetint}) be flat. We suppose that the initial system
(\ref{abe:DNriem}) is a DN system which possesses Riemann
invariants and we let $g_{ii}(\vu)$ be one of the metrics
associated to it. Under such hypotheses, we prove that if the
reciprocal metric ${\hat g}_{ii} (\vu)$ in (\ref{abe:recmetint})
is flat, then the reciprocal transformation is canonical for the
initial metric $g_{ii}(\vu)$.

Then, we restrict ourselves to the case in which the initial
metric is either flat or of constant curvature and, in Theorem
\ref{abe:teosuff}, we classify the reciprocal transformations of
one or both the independent variables so that the reciprocal
metric be flat. Finally, when both the initial and the transformed
metrics are flat, we also discuss the geometric intepretation of
the latter Theorem in view of the results obtained by Ferapontov
in \cite{abe:F95-1}. Indeed, in Theorem \ref{abe:cor} we show
that, the hypersurfaces of two diagonalizable DN systems are Lie
equivalent if and only if the corresponding local hamiltonian
structures are related by a canonical reciprocal transformation
which satisfies Theorem \ref{abe:teosuff}.

There are of course still many open problems connected to the
classification of local Hamiltonian structures: what about the
possible role of other types of transformations among hydrodynamic
type systems?  What is the geometrical meaning of the conditions
settled by Theorems \ref{abe:teonec1}, \ref{abe:teonec2} and
\ref{abe:teosuff} when the initial metric is not flat? Moreover,
there do exists non--diagonalizable integrable Hamiltonian
systems; it would be interesting to check whether the same
conditions on the conservation laws in the reciprocal
transformations preserving the locality of the Hamiltonian
structure still hold also in that case.

Finally, several systems of evolutionary PDEs arising in physics
may be written as perturbations of hyperbolic systems of PDEs and
their classification in case of Hamiltonian perturbations has
recently been started by Dubrovin, Liu and Zhang~\cite{abe:DLZ06}.
It would also be interesting to investigate the role of reciprocal
transformations in this perturbation scheme.

The plan of the paper is as follows. In the next section, we
introduce the necessary definitions and we recall some theorems we
proved in~\cite{abe:AG07} on the form of the reciprocal Riemannian
curvature tensor and of the reciprocal Hamiltonian operator. In
section 3, we prove the necessary conditions on the form of the
Riemannian curvature tensor and the conservation laws in the
reciprocal transformation so that the reciprocal metric be flat.
Finally in section 4, we classify the reciprocal transformation
which preserve the flatness of the metric or which transform a
constant curvature metric to a flat one and we apply such
conditions to some examples.

\section{The reciprocal Hamiltonian structure}
In this section we introduce some useful notations, we discuss the
role of additive constants in the extended reciprocal
transformations and we recall some theorems we proved in
\cite{abe:AG07} which we shall use in the following sections.

We consider a smooth DN Hamiltonian hydrodynamic system in Riemann
invariants
\begin{equation}\label{abe:sisdia}
u^i_t =v^i(\vu)u^i_x, \quad\quad i=1,\dots, n,
\end{equation}
with $v^i(\vu)$ all real and distinct (strict hyperbolicity
property). Let $g_{ii} (\vu)$ be a (covariant) non--degenerate
diagonal metric such that for convenient $f^i(u^i)$, $i=1,\dots,
n$, $g_{ii} (\vu)f^i(u^i)$ is a flat metric associated to the
local Hamiltonian operator of the system (\ref{abe:sisdia}). Let
$g^{ii} (\vu) = 1/g_{ii} (\vu)$. Let $H_i (\vu)$,
$\beta_{ij}(\vu)$ and ${ \Gamma}^{i}_{jk}(\vu)$ respectively be
the Lam\'e coefficients the rotation coefficients and the
Christoffel symbol of a diagonal non-degenerate metric $g_{ii}
(\vu)$ associated to (\ref{abe:sisdia}),
\[
H_i (\vu) = \sqrt{g_{ii} (\vu)},\quad\quad \beta_{ij} (\vu) =
\frac{\partial_i H_j (\vu)}{H_i(\vu)}, \quad i\not = j,
\]
\[
{ \Gamma}^{i}_{jk}(\vu)= \frac{1}{2} g^{im} (\vu) \left(
\frac{\partial  g_{mk}(\vu)}{\partial u^j} + \frac{\partial
g_{mj}(\vu)}{\partial u^k} -\frac{\partial g_{kj}(\vu)}{\partial
u^m} \right).
\]
Since the metric is diagonal, the only non--zero Christoffel
symbols are
\[
\begin{array}{ll}
{ \Gamma}^{j}_{ii}(\vu) &=\displaystyle
-\frac{H_i(\vu)}{H_j^2(\vu)}
\partial_j H_i (\vu), \quad\quad \forall i\not =j,\\
&\\ { \Gamma}^{i}_{ij}(\vu) &=\displaystyle \frac{\partial_j
H_i(\vu)}{H_i(\vu)},\quad\quad \forall i,j=1,\dots,n.
\end{array}
\]
Under our hypotheses, the system (\ref{abe:sisdia}) possesses at
least one flat metric. Then, for any other metric associated to
(\ref{abe:sisdia}), the Euler--Darboux equations
(\ref{abe:darboux2}) still hold,
\[
\partial_k
\beta_{ij}(\vu)-\beta_{ik}(\vu)\beta_{kj}(\vu)\equiv 0,\quad\quad
i\not = j\neq k,\] that is $R^{ij}_{ik} (\vu) \equiv 0$,
($i\not=j\not=k\not= i$). For systems (\ref{abe:sisdia}),
Ferapontov \cite{abe:F95} constructs the non--local Hamiltonian
operators $J^{ij} (\vu)$ associated to non-flat metrics $g_{ii}
(\vu)$ which take the form
\begin{equation}\label{abe:ham}
J^{ij} (\vu) =g^{ii} (\vu) \left(\delta^i_j \frac{d}{dx}-
\Gamma^{j}_{ik} (\vu) u^k_x\right)+\sum_{l} \epsilon^{(l)}
w^i_{(l)} (\vu) u^i_x \left( \frac{d}{dx} \right)^{-1} w^j_{(l)}
(\vu) u^j_x,
\end{equation}
where $\epsilon^{l}=\pm 1$, $w^{i}_{(l)} (\vu)$ are affinors of
the metric which satisfy
\begin{equation}\label{abe:commuting}
\frac{\partial_j w^i_{(l)}(\vu)}{w^j_{(l)}(\vu)-w^i_{(l)}(\vu)}
=\frac{\partial_j v^i(\vu)}{v^j(\vu)-v^i(\vu)}= \partial_j \ln H_i
(\vu),
\end{equation}
and the curvature tensor of the metric takes the form
\begin{equation}\label{abe:curvik}
\displaystyle R^{ik}_{ik} (\vu) = -\frac{\Delta_{ik}(\vu)}{H_i
(\vu) H_k(\vu)} \equiv \sum_{(l)} \epsilon^{l} w^{i}_{(l)} (\vu)
w^{k}_{(l)} (\vu) ,\quad\quad i\not = k,
\end{equation}
where
\[
\Delta_{ik}(\vu)=\partial_i \beta_{ik} (\vu)+\partial_k
\beta_{ki}(\vu) +\sum_{m\not = i,k} \beta_{mi}
(\vu)\beta_{mk}(\vu).
\]
\begin{remark}\label{abe:poisson}
In particular, if $g_{ii} (\vu)$ is flat, then $\displaystyle
J^{ij} (\vu) =g^{ii} (\vu) \left(\delta^i_j \frac{d}{dx}-
\Gamma^{j}_{ik} (\vu) u^k_x\right)$ \cite{abe:DN83}.

\noindent If $g_{ii} (\vu)$ is of constant curvature $c$, then
\cite{abe:FM90}
\begin{equation}\label{abe:Poiconst}
J^{ij} (\vu) = g^{ii} (\vu) \left(\delta^i_j \frac{d}{dx}-
\Gamma^{j}_{ik} (\vu) u^k_x\right)+c u^i_x \left( \frac{d}{dx}
\right)^{-1}  u^j_x.
\end{equation}
If $g_{ii} (\vu)$ is conformally flat, then
\begin{equation}\label{abe:Rij}
R^{ij}_{ij} (\vu) = w^i(\vu) + w^j(\vu), \quad i\not = j,
\end{equation}
and \begin{equation}\label{abe:Poiconf} J^{ij} (\vu) = g^{ii}
(\vu) \left(\delta^i_j \frac{d}{dx}- \Gamma^{j}_{ik} (\vu)
u^k_x\right)+ w^i (\vu) u^i_x \left( \frac{d}{dx} \right)^{-1}
u^j_x + u^i_x \left( \frac{d}{dx} \right)^{-1}  w^j (\vu)u^j_x.
\end{equation}
In the next section a special role is played by the metrics
$g_{ii}(\vu)$ for which the Riemannian curvature tensor takes the
special form
\begin{equation}\label{abe:curvspe}
R^{ik}_{ik} (\vu)= w^i_{(1)} (\vu) + w^k_{(1)}  (\vu) + w^i_{(2)}
(\vu) v^{k} (\vu) + w^k_{(2)} (\vu) v^{i} (\vu),\quad\quad i\not
=k,
\end{equation}
and
\begin{equation}\label{abe:Poispe}
\begin{array}{rl}
\displaystyle J^{ij} (\vu) &= \displaystyle g^{ii} (\vu)
\left(\delta^i_j \frac{d}{dx}- \Gamma^{j}_{ik} (\vu) u^k_x\right)+
w^i_{(1)} (\vu) u^i_x \left( \frac{d}{dx} \right)^{-1} u^j_x +
u^i_x \left( \frac{d}{dx} \right)^{-1}  w^j_{(1)} (\vu)u^j_x\\
&\displaystyle + w^i_{(2)} (\vu) u^i_x \left( \frac{d}{dx}
\right)^{-1} v^j (\vu) u^j_x + v^i (\vu)u^i_x \left( \frac{d}{dx}
\right)^{-1} w^j_{(2)} (\vu)u^j_x
\end{array}
\end{equation}
\end{remark}
Given smooth conservation laws
\[
B(\vu)_t=A(\vu)_x,\quad N(\vu)_t=M(\vu)_x
\]
for the system (\ref{abe:sisdia}), a reciprocal transformation of
the independent variables $x,t$ is defined by \cite{abe:RS82}
\begin{equation}\label{abe:xt}
d{\hat x} = B (\vu)dx + A(\vu)dt,\quad\quad d{\hat t} = N(\vu)dx
+M(\vu)dt.
\end{equation}
In \cite{abe:FP03}, Ferapontov and Pavlov have characterized the
tensor of the reciprocal Riemannian curvature and the reciprocal
Hamiltonian structure when the initial metric $g_{ii}(\vu)$ is
flat. In \cite{abe:AG07}, we have computed the Riemannian
curvature and the Hamiltonian structure of the reciprocal system
\begin{equation}\label{abe:sistfin}
u^{i}_{{\hat t}} = {\hat v}^i (\vu) u^{i}_{{\hat x}}
=\frac{B(\vu)v^i(\vu)-A(\vu)}{M(\vu)-N(\vu)v^i(\vu)}u^{i}_{{\hat
x}},
\end{equation}
associated to the reciprocal metric
\begin{equation}\label{abe:tramet}
{\hat g}_{ii}(\vu)=
\left(\frac{M(\vu)-N(\vu)v^i(\vu)}{B(\vu)M(\vu)-A(\vu)N(\vu)}
\right)^2{ g}_{ii}(\vu),
\end{equation}
with $g_{ii} (\vu)$ non-flat. In the following, we use the symbols
${\hat H}_i(\vu)$, ${\hat \beta}_{ij}(\vu)$, ${
\hat{\Gamma}}^{i}_{jk}(\vu)$, ${\hat R}^{ij}_{km}(\vu)$ and
$\hat{J}^{ij}$, respectively, for the Lam\'e coefficients, the
rotation coefficients, the Christoffel symbols, the Riemannian
curvature tensor and the Hamiltonian operator associated the
reciprocal metric ${\hat g}_{ii}(\vu)$. To simplify notations, we
drop the $\vu$ dependence in the lengthy formulas.

\begin{theorem}\label{abe:theo3.1}\cite{abe:AG07}
Let ${ g}_{ii} (\vu)$ be the covariant diagonal metric as above
for the Hamiltonian system (\ref{abe:sisdia}) with Riemannian
curvature tensor as in (\ref{abe:curvik}) or as in
(\ref{abe:curvspe}). Then, for the reciprocal metric ${\hat
g}_{ii}(\vu)$ defined in (\ref{abe:tramet}), the only possible
non-zero components of the reciprocal Riemannian curvature tensor
are
\begin{equation}\label{abe:tracurvd}
\begin{array}{rcl}
\displaystyle {\hat R}^{ik}_{ik} (\vu)&=&\displaystyle \frac{{
H}_i{ H}_k}{{\hat H}_i{\hat H}_k}{ R}^{ik}_{ik} -({ \nabla} B)^2+
\frac{{ H}_k}{{\hat H}_k}{ \nabla}^k{ \nabla}_k B+\frac{{
H}_i}{{\hat H}_i}{ \nabla}^i{ \nabla}_i B-{\hat
v}^k{\hat v}^i({ \nabla} N)^2\\
&&\\
\displaystyle &&\displaystyle +{\hat v}^k\frac{{ H}_i}{{\hat
H}_i}{ \nabla}^i{ \nabla}_i N+{\hat v}^i\frac{{ H}_k}{{\hat H}_k}{
\nabla}^k{ \nabla}_k N- ({\hat v}^k+{\hat v}^i)<{ \nabla} B,{
\nabla} N>,\quad i\not=k\\
\end{array}
\end{equation}
where
\[
<{ \nabla} B(\vu),{ \nabla} N(\vu)> = \sum_{m} { g}^{mm}(\vu)
\partial_m B (\vu)\,
\partial_m N(\vu),\]
\[
{ \nabla}^i{ \nabla}_i B(\vu) = { g}^{ii} (\vu)\left(
\partial_i^2 B (\vu)-\sum_{m} { \Gamma}^m_{ii}(\vu)\,
\partial_m B (\vu)\right),
\]
\[
{ \nabla}^i{ \nabla}_j B (\vu)= { g}^{ii} (\vu)\left(
\partial_i\partial_j B (\vu)-{ \Gamma}^i_{ij}(\vu)\,
\partial_i B(\vu) -{ \Gamma}^j_{ij}(\vu)\,
\partial_j B(\vu)\right).
\]
\end{theorem}

\medskip

In \cite{abe:AG07}, we computed the reciprocal affinors and the
reciprocal Hamiltonian operator of a hydrodynamic system
(\ref{abe:sisdia}) with (nonlocal) Hamiltonian operator
(\ref{abe:ham}). At this aim, we introduce the auxiliary flows
\begin{equation}\label{abe:bnflow}
u^i_{\tau} = n^{i} (\vu) u^i_x = { J}^{ij} (\vu)
\partial_j N(\vu),
\quad\quad \displaystyle u^i_{\zeta} = b^{i} (\vu) u^i_x = {
J}^{ij} (\vu)
\partial_j B(\vu),
\end{equation}
\[
u^i_{t_{(l)}} = w^i_{(l)}(\vu) u^i_x = { J}^{ij} (\vu)
\partial_j H^{(l)}(\vu),\]
respectively, generated by the densities of conservation laws
associated to the reciprocal transformation (\ref{abe:xt}),
$B(\vu)$, $N(\vu)$, and by the densities of conservation laws
$H^{(l)}(\vu)$ associated to the affinors $w^i_{(l)}(\vu)$ of the
Riemannian curvature tensor (\ref{abe:curvik}). By construction,
all the auxiliary flows commute with (\ref{abe:sisdia}).
Introducing the following closed form
\begin{equation}\label{abe:clofor}
\left\{ \begin{array}{l} \displaystyle d{\hat x} = B (\vu)dx +
A(\vu)dt+ P(\vu)d\tau + Q(\vu) d\zeta + \sum_{l} T^{(l)}
(\vu)dt_{(l)},
\\\displaystyle d{\hat t} = N(\vu)dx + M(\vu)dt+R(\vu)d\tau+
S(\vu)d\zeta + \sum_{l} Z^{(l)}(\vu)
dt_{(l)},\\
\displaystyle d{\hat \tau} = d\tau, \quad d{\hat \zeta} = d\zeta,
\quad d{\hat t}_{(l)} = dt_{(l)},
\end{array}\right.\end{equation}
it is easy to verify that the reciprocal auxiliary flows
\[
u^i_{{\hat \tau}} = {\hat n}^i(\vu) u^i_{{\hat x}}, \quad\quad
u^i_{{\hat \zeta}} = {\hat b}^i(\vu) u^i_{{\hat x}}, \quad\quad
u^i_{{\hat t}^{(l)}} = {\hat w}^i_{(l)} (\vu) u^i_{{\hat x}},
\]
satisfy
\begin{equation}\label{abe:travel}
\begin{array}{l}
\displaystyle {\hat n}^i(\vu) =  n^i B - P + (N n^i-R){\hat v}^i
 = \frac{{ H}_i}{{\hat H}_i}n^i-P-{\hat v}^iR
,
\\
\displaystyle {\hat b}^i(\vu) = b^i B - Q + (Nb^i-S){\hat v}^i
 = \frac{{ H}_i}{{\hat H}_i}b^i-Q-{\hat v}^iS,
\\
\displaystyle {\hat w}^i_{(l)}(\vu) =  w^i_{(l)} B - T^{(l)} +
(Nw^i_{(l)}-Z^{(l)}){\hat v}^i = \frac{{ H}_i}{{\hat
H}_i}w^i_{(l)}-T^{(l)}-{\hat v}^iZ^{(l)} .
\end{array}
\end{equation}
Using (\ref{abe:clofor}), we immediately conclude that
$T^{(l)}(\vu)$, $Z^{(l)}(\vu)$ satisfy
\begin{equation}\label{abe:bnflow1}
n^{i} (\vu) = { \nabla}^i{ \nabla}_i N + \sum_{(l)} \epsilon_{(l)}
Z^{(l)} w^i_{(l)},\quad b^{i} (\vu) = { \nabla}^i{ \nabla}_i B +
\sum_{(l)} \epsilon_{(l)} T^{(l)} w^i_{(l)}.
\end{equation}
Moreover, we have
\begin{equation}\label{abe:clofl}
\begin{array}{l}
\displaystyle v^i(\vu) = \frac{\partial_i A(\vu)}{\partial_i
B(\vu)}= \frac{\partial_i M(\vu)}{\partial_i N(\vu)}, \quad\quad
w^i_{(l)}(\vu) = \frac{\partial_i T^{(l)} (\vu)}{\partial_i
B(\vu)}= \frac{\partial_i Z^{(l)}(\vu)}{\partial_i N(\vu)},\\
\displaystyle b^i(\vu)  = \frac{\partial_i Q(\vu)}{\partial_i
B(\vu)}= \frac{\partial_i S(\vu)}{\partial_i N(\vu)},\quad\quad
n^i(\vu)= \frac{\partial_i P(\vu)}{\partial_i B(\vu)}=
\frac{\partial_i R(\vu)}{\partial_i N(\vu)}.
\end{array}
\end{equation}
Using (\ref{abe:bnflow1}) and (\ref{abe:clofl}), $Q(\vu)$,
$R(\vu)$ and $P(\vu)+S(\vu)$ are uniquely defined (up to additive
constants) by the following identities
\begin{equation}\label{abe:grads}
\begin{array}{c}
\displaystyle Q(\vu) =\frac{1}{2}\left( { \nabla} B \right)^2 +
\frac{1}{2}\sum_{l}\epsilon_{(l)}\left( T^{(l)} \right) ^2,
\quad\quad R(\vu) =\frac{1}{2}\left( { \nabla} N \right)^2
+\frac{1}{2}\sum_{l}\epsilon_{(l)}\left( Z^{(l)} \right) ^2,
\\
\displaystyle P(\vu)+S(\vu) = < { \nabla} N, { \nabla} B > +
\sum_{l}\epsilon_{(l)}T^{(l)} Z^{(l)}.
\end{array}
\end{equation}
If the Riemannian curvature tensor associated to $g_{ii} (\vu)$
takes the special form (\ref{abe:curvspe}), then
(\ref{abe:bnflow1}) take the special form
\begin{equation}\label{abe:bnflowspe}
\begin{array}{c}
n^{i} (\vu) = { \nabla}^i{ \nabla}_i N + w^i_{(1)} N+Z^{(1)} +
w^i_{(2)} M+ v^i Z^{(2)},\\
b^{i} (\vu) = { \nabla}^i{ \nabla}_i B + w^i_{(1)} B+T^{(1)} +
w^i_{(2)} A+ v^i T^{(2)}, \end{array}
\end{equation}
and
\begin{equation}\label{abe:gradsspe}
\begin{array}{c}
\displaystyle Q(\vu) =\frac{1}{2} (\nabla B)^2 (\vu) + B(\vu)
T^{(1)} (\vu)+ A(\vu)  T^{(2)} (\vu),\\\displaystyle  R(\vu)
=\frac{1}{2} (\nabla N)^2 (\vu) + N(\vu) Z^{(1)} (\vu)+ M(\vu)
Z^{(2)} (\vu),
\\
\displaystyle P(\vu)+S(\vu)=< { \nabla} N, { \nabla} B > +T^{(1)}
N+ T^{(2)} M+ Z^{(1)} B +Z^{(2)} A.
\end{array}
\end{equation}

\begin{remark}\label{abe:remark1}
The addition of constants to the r.h.s. of (\ref{abe:grads}) leave
invariant the reciprocal transformation is the sense that the
reciprocal metric ${\hat g}_{ii}(\vu)$, the reciprocal Riemannian
tensor ${\hat R}^{ij}_{ij} (\vu)$, the reciprocal Hamiltonian
operator ${\hat J}^{ij} (\vu)$ and the reciprocal Hamiltonian
velocity flow ${\hat v}^i (\vu)$ are not effected by them. These
constants just effect the auxiliary flows. Indeed, let $Q(\vu)$,
$P(\vu)$, $R(\vu)$ and $S(\vu)$ be as in (\ref{abe:grads}) and let
us consider the modified closed form
\[
\left\{ \begin{array}{l} \displaystyle d{\hat x} = B (\vu)dx +
A(\vu)dt+ (P(\vu)+\alpha) d\tau + (Q(\vu)+\beta) d\zeta + \sum_{l}
T^{(l)} (\vu)dt_{(l)},
\\\displaystyle d{\hat t} = N(\vu)dx + M(\vu)dt+(R(\vu)+\gamma)d\tau+
(S(\vu)+\delta) d\zeta + \sum_{l} Z^{(l)}
(\vu)dt_{(l)},\\
\displaystyle d{\hat \tau} = d\tau, \quad d{\hat \zeta} = d\zeta,
\end{array}\right.
\]
with $\alpha,\beta,\gamma,\delta$ arbitrary constants.
\[
{\hat n}^i_m(\vu) = {\hat n}^i(\vu) -\beta -\delta{\hat v}^i(\vu),
\quad\quad {\hat b}^i_m (\vu) = {\hat b}^i (\vu) -\alpha
-\gamma{\hat v}^i(\vu),
\]
with ${\hat n}^i(\vu)$ and ${\hat b}^i (\vu)$ as
in(\ref{abe:travel}).
\end{remark}

The following alternative expressions for the reciprocal Riemann
curvature tensor and the reciprocal Hamiltonian structure hold.
\begin{theorem}\label{abe:theo3.2}
Let $g_{ii} (\vu)$ be the metric for the Hamiltonian hydrodynamic
system (\ref{abe:sisdia}), with Riemannian curvature tensor as in
(\ref{abe:curvik}). Then, after the reciprocal transformation
(\ref{abe:xt}), the non zero components of the reciprocal
Riemannian curvature tensor are
\begin{equation}\label{abe:nonimp}
{\hat R}^{ik}_{ik} (\vu) = \sum_{l}\epsilon^{(l)} {\hat
w}^i_{(l)}(\vu) {\hat w}^k_{(l)}(\vu) + {\hat v}^i(\vu) {\hat
n}^k(\vu) +{\hat v}^k (\vu){\hat n}^i(\vu) + {\hat b}^i (\vu)+
{\hat b}^k(\vu), \quad\quad i\not = k,
\end{equation}
where the reciprocal metric ${\hat g}_{ii}(\vu)$ and the
reciprocal affinors ${\hat n}^i(\vu)$, ${\hat b}^i(\vu)$ and
${\hat w}^i_{(l)}(\vu)$ are as in (\ref{abe:tramet}) and
(\ref{abe:travel}), respectively, with $Q(\vu)$, $P(\vu)$,
$R(\vu)$ and $S(\vu)$ as in (\ref{abe:grads}).

Let $g_{ii} (\vu)$ be the metric for the Hamiltonian hydrodynamic
system (\ref{abe:sisdia}), with Riemannian curvature tensor as in
(\ref{abe:curvspe}), then the nonzero components of the
transformed curvature tensor take the form
\begin{equation}\label{abe:curvnostro}  {\hat R}^{ik}_{ik} (\vu) =
{\hat n}^i(\vu) {\hat v}^k(\vu) + {\hat n}^k(\vu) {\hat v}^i(\vu)
+ {\hat b}^i (\vu)+ {\hat b}^k(\vu), \quad\quad i\not = k,
\end{equation} where the reciprocal metric ${\hat g}_{ii}(\vu)$
and the reciprocal affinors ${\hat
n}^i(\vu)$, ${\hat b}^i(\vu)$ and ${\hat w}^i_{(l)}(\vu)$ are as
in (\ref{abe:tramet}) and (\ref{abe:travel}), respectively, with
$Q(\vu)$, $P(\vu)$, $R(\vu)$ and $S(\vu)$ as in
(\ref{abe:gradsspe}).
\end{theorem}

Formula (\ref{abe:nonimp}) has already been proven in
\cite{abe:AG07}. To prove (\ref{abe:curvnostro}), it is sufficient
to insert (\ref{abe:bnflowspe}) and (\ref{abe:gradsspe}) into
(\ref{abe:tracurvd}).

\begin{corollary}\label{abe:corxt}
Let the reciprocal transformation changes only $x$ ($N(\vu)=0$ and
$M(\vu)=1$ in (\ref{abe:xt})), then the nonzero components of the
transformed curvature tensor take the form
\begin{equation}\label{abe:curvx}
\displaystyle {\hat R}^{ik}_{ik} (\vu) =\; \displaystyle B^2(\vu)
{ R}^{ik}_{ik} (\vu)+ B (\vu)({ \nabla}^i{ \nabla}_i B(\vu)+ {
\nabla}^k{ \nabla}_k B(\vu)) - ({ \nabla} B(\vu))^2
\end{equation}
Moreover, if the Riemannian curvature tensor of $g_{ii} (\vu)$
takes the form as in (\ref{abe:curvik}), then
\[
{\hat R}^{ik}_{ik} (\vu) =\displaystyle \sum_{l}\epsilon^{(l)}
{\hat w}^i_{(l)}(\vu) {\hat w}^k_{(l)}(\vu) + {\hat b}^i (\vu)+
{\hat b}^k(\vu);
\]
if Riemannian curvature tensor associated to $g_{ii} (\vu)$ takes
the form (\ref{abe:curvspe}) then the nonzero components of the
transformed curvature tensor take the form
\begin{equation}\label{abe:curvx1}
\displaystyle {\hat R}^{ik}_{ik} (\vu) =\;  {\hat w}^i_{(2)}(\vu)
{\hat v}^k_{(l)}(\vu) + {\hat w}^k_{(2)}(\vu) {\hat
v}^i_{(l)}(\vu) + {\hat b}^i (\vu)+ {\hat b}^k(\vu).
\end{equation}

\smallskip

If the reciprocal transformation changes only $t$ ($B(\vu)=1$ and
$A(\vu)=0$ in (\ref{abe:xt})), then the nonzero components of the
transformed curvature tensor satisfy
\begin{equation}\label{abe:curvt}
\displaystyle {\hat R}^{ik}_{ik} (\vu) = \;\displaystyle\frac{M^2
{ R}^{ik}_{ik} + M \, (v^k\, { \nabla}^i{ \nabla}_i N+ v^i\, {
\nabla}^k{ \nabla}_k N) - v^i\, v^k\, ({ \nabla} N)^2}{(M-Nv^i)
(M-Nv^k)}
\end{equation}
Moreover, if the Riemannian curvature tensor of $g_{ii} (\vu)$
takes the form as in (\ref{abe:curvik}), then
\[
{\hat R}^{ik}_{ik} (\vu) = \displaystyle \;\sum_{l}\epsilon^{(l)}
{\hat w}^i_{(l)}(\vu){\hat w}^k_{(l)}(\vu)+ {\hat v}^i(\vu) {\hat
n}^k(\vu) +{\hat v}^k(\vu) {\hat n}^i(\vu);
\]
if Riemannian curvature tensor associated to $g_{ii} (\vu)$ takes
the form (\ref{abe:curvspe}), then the nonzero components of the
transformed curvature tensor take the form
\begin{equation}\label{abe:curvt1}
\displaystyle {\hat R}^{ik}_{ik} (\vu) = \;{\hat
w}^i_{(1)}(\vu)+{\hat w}^k_{(1)}(\vu)+ {\hat v}^i(\vu) {\hat
n}^k(\vu) +{\hat v}^k(\vu) {\hat n}^i(\vu).
\end{equation}
\end{corollary}

Formulas (\ref{abe:curvx}), (\ref{abe:curvt}) and their
expressions when $R^{ik}_{ik} (\vu)$ is as in (\ref{abe:curvik})
have already been proven in \cite{abe:AG07}. To prove
(\ref{abe:curvx1}) (resp. (\ref{abe:curvt1})) it is sufficient to
insert (\ref{abe:bnflowspe}) and (\ref{abe:gradsspe}) into
(\ref{abe:curvx}) (resp. (\ref{abe:curvt})).

\section{Necessary conditions for reciprocal flat metrics}

In this section, we start from an integrable Hamiltonian system
$u^i_t = v^i (\vu) u^i_x$, $i=1,\dots,n$ and we investigate the
necessary conditions on the initial metric and on the conservation
laws in the reciprocal transformation so that the reciprocal
metric be flat. The conditions settled by Theorem
\ref{abe:teonec2} on the conservation laws in the reciprocal
transformations are very strict: if $n\ge 5$, they must be linear
combinations with constant coefficients of the Casimirs, the
momentum and the Hamiltonian densities with respect to the initial
Hamiltonian structure. The same Theorem settles also very strict
conditions on the admissible form of the Riemannian curvature
tensor associated to the initial metric $g_{ii}(\vu)$. In the case
of reciprocal transformations of a single independent variable the
necessary conditions are even more restrictive: if $n\ge 3$, the
conservation law is a linear combination of Casimirs and momentum
densities (respectively of Casimirs and Hamiltonian densities) if
just the $x$ variable (resp. the $t$ variable) changes.

\begin{definition}
Following Ferapontov \cite{abe:F95-1,abe:F95}, we call canonical a
reciprocal transformation as in (\ref{abe:xt}), in which the
integrals, up to additive constants, are linear combinations of
the canonical integrals (Casimirs, Hamiltonian and momentum) with
respect to the given Hamiltonian structure.
\end{definition}

\begin{remark}
If the initial metric $g_{ii} (\vu)$ is not flat, a Casimir
density (resp. a momentum density, a Hamiltonian density)
associated to the corresponding non-local Hamiltonian operator
$J^{ij} (\vu)$ in (\ref{abe:ham}) is a conservation law $h(u)$
such that $J^{ij}
\partial_j h(\vu)\equiv 0$ (resp. $J^{ij} \partial_j h(\vu)\equiv
u^i_x$, $J^{ij} \partial_j h(\vu)\equiv v^i(\vu) u^i_x$). We
remark that, under the hypotheses of the following Theorem, for
each Hamiltonian structure with $k$ non--localities in the
Hamiltonian operator, there do exist $(n+k+2)$ canonical integrals
as proven by Maltsev and Novikov\cite{abe:NM}.
\end{remark}

\smallskip

In the following Theorem we settle the necessary conditions for
reciprocal flat metrics in the case of a transformation of a
single variable.

\begin{theorem}\label{abe:teonec1}  Let $u^i_t =v^i(\vu)u^i_x$,
$i=1,\dots,n$, $n\ge 3$, be
an integrable strictly hyperbolic DN hydrodynamic type system as
in (\ref{abe:sisdia}), let ${ g}_{ii}(\vu)$ be one of its metrics
with Hamiltonian operator $J^{ij} (\vu)$ as in (\ref{abe:ham}).

\smallskip

i) Let $d{\hat x} = B(\vu) d{ x} + A (\vu) d{t}$, $d{\hat t} = dt
$, be a reciprocal transformation such that the reciprocal metric
${\hat g}_{ii} (\vu)$ defined in (\ref{abe:tramet}) be flat.

Then $B (\vu)$ is a linear combination of the Casimirs and the
momentum densities (up to an additive constant), and $g_{ii}(\vu)$
is either a flat or a constant curvature or a conformally flat
metric.

\smallskip

ii) Let $d{\hat x} = d{ x}$, $d{\hat t} = N (\vu) d x + M (\vu) d
t$, be a reciprocal transformation such that the reciprocal metric
${\hat g}_{ii} (\vu)$ defined in (\ref{abe:tramet}) be flat. In
the case $n=3$, let moreover $v^i (\vu) \not \equiv 0$,
$i=1,\dots,3$.

Then $N (\vu)$ is a linear combination of the Casimirs and the
Hamiltonian densities (up to an additive constant), and the
Riemannian curvature tensor associated to the initial metric
$g_{ii}(\vu)$ takes the form
\begin{equation}\label{abe:evid}
R^{ij}_{ij} (\vu) =w^i (\vu) v^{j} (\vu) + w^j (\vu) v^{i} (\vu)
,\quad\quad i\not =j,\end{equation} with $w^i (\vu)$ (possibly
null) affinors.
\end{theorem}

\smallskip

{\sl Proof of the theorem} To compute the form of the Riemannian
curvature tensor associated to the initial metric $g_{ii}(\vu) $
it is sufficient to invert the reciprocal transformation
(\ref{abe:dirtrasf}) and to apply Theorem \ref{abe:theo3.2} to the
reciprocal flat metric ${\hat g}_{ii} (\vu)$.

{\it i)} If the reciprocal transformation changes only $x$
($N(\vu)\equiv 0$, $M(\vu)\equiv 1$) and the reciprocal metric
${\hat g}_{ii} (\vu)$ is flat, the Riemann curvature tensor
associated to the initial metric $g_{ii} (\vu)$ takes the form
$R^{ik}_{ik} (\vu)=w^i_{(1)}(\vu)+w^k_{(1)}(\vu)$, ($i\not =k$),
with possibly constant or null affinors $w^i_{(1)}(\vu)$ (see
\cite{abe:FP03}). According to Corollary (\ref{abe:corxt}), the
zero curvature equations ${\hat R}^{ik}_{ik} (\vu)={\hat b}^i
(\vu)+{\hat b}^k (\vu)\equiv 0$, ($i\not = k$), for the reciprocal
metric ${\hat g}_{ii} (\vu)$ are then equivalent to
\[
0\equiv {\hat b}^i (\vu) = B(\vu) b^i(\vu) -Q(\vu), \quad\quad
i=1,\dots,n, \] as follows from (\ref{abe:curvx1}) with $Q(\vu)$
as in (\ref{abe:gradsspe}). Since $b^{i} (\vu) =\frac{\partial_i
Q(\vu)}{\partial_i B(\vu)}$, ($i=1,\dots, n$), we immediately
conclude that there exists a constant $\kappa$ such that
\[
u^i_{{\hat\zeta}} \equiv b^i (\vu) u^i_{x} \equiv J^{ij}
(\vu)\partial_j  B(\vu)= \kappa u^i_x, \quad\quad i=1,\dots,n, \]
that is $B(\vu)$ is a linear combination of the Casimirs and the
momentum densities up to an additive constant.

\smallskip

{\it ii)} Similarly, if the reciprocal transformation changes only
$t$ ($B(\vu)\equiv 1$, $A(\vu)\equiv 0$) and the reciprocal metric
${\hat g}_{ii} (\vu)$ is flat, the Riemann curvature tensor
associated to the initial metric $g_{ii} (\vu)$ takes the form
$R^{ik}_{ik} (\vu) = w^i_{(2)}(\vu) v^k(\vu) +
w^k_{(2)}(\vu)v^i(\vu)$, ($i\not =k$), with possibly constant or
null affinors $w^i_{(2)}(\vu)$ (see \cite{abe:FP03}). According to
Corollary (\ref{abe:corxt}), the zero curvature equations for the
reciprocal metric, ${\hat R}^{ik}_{ik} (\vu)={\hat v}^i(\vu) {\hat
n}^k(\vu) +{\hat v}^k(\vu) {\hat n}^i(\vu)\equiv 0$, ($i\not =
k$), are equivalent to \begin{equation}\label{abe:casodeli}
0\equiv {\hat n}^i (\vu) = \frac{M(\vu) n^i(\vu) - R(\vu)
v^i(\vu)}{M(\vu) -N(\vu) v^i(\vu)}, \quad\quad
i=1,\dots,n.\end{equation} Since $v^{i} (\vu) =\frac{\partial_i
M(\vu)}{\partial_i N(\vu)}$, $n^{i} (\vu) =\frac{\partial_i
R(\vu)}{\partial_i N(\vu)}$, ($i=1,\dots, n$), we immediately
conclude that there exists a constant $\kappa$ such that
\[
u^i_{{\hat\tau}} \equiv n^i (\vu) u^i_{x} \equiv J^{ij} (\vu)
\partial_j  N(\vu)= \kappa v^i (\vu)  u^i_{x}, \quad\quad
i=1,\dots,n,\] that is the density of conservation law associated
to the inverse transformation is a linear combination of the
Casimirs and the Hamiltonian densities up to an additive constant.
$\quad \square$

\medskip

\begin{remark}
The Theorem \ref{abe:teonec1} is not applicable in the case $n=2$.
For instance, in the case of a transformation of the single
variable $x$, we get the zero curvature condition ${\hat b}^1(\vu)
=- {\hat b}^2(\vu)$ and it is possible to construct non-canonical
reciprocal transformations which preserve the flatness of the
metric. Here is a counterexample suggested by the second referee:
let us take a linear 2-component system
\[
u^1_t=pu^1_x, \quad\quad u^2_t=qu^2_x,
\]
where $p, q$ are constant. It has infinitely many Hamiltonian
structures, let's take the one corresponding to the metric $g =
(du^1)^2 + (du^2)^2$. Let us consider a reciprocal transformation
of $x$ only, $d{\hat x}= B(u^1,u^2) dx + A(u^1,u^2) dt$, ${\hat
t}=t$. For the above system, the general form of a density of
conservation law is $B(u^1,u^2)=f^1(u^1)+ f^2(u^2)$. Let us
require that the transformed metric be flat: this gives a
functional-differential equation for $f^1$ and $f^2$ which can be
solved explicitly.

In particular, if $\displaystyle
B(u^1,u^2)=a+bu^1+cu^2+\frac{d}{2}((u^1)^2+(u^2)^2)$, then the
flatness condition gives $b^2+c^2=2ad$. This is the case of
canonical integrals discussed in Theorem \ref{abe:teonec1}.
However, there is another solution:
\[
B(u^1,u^2)= a\exp( u^1) + b\exp (-u^1) + c \sin (u^2) + d \cos
(u^2)
\]
with $c^2+d^2=4ab$. Thus, the reciprocal metric is flat, although
the density $B$ is not a linear combination of canonical
integrals.
\end{remark}

\medskip

\begin{remark}
In the case of time transformations and $n=3$, the hypothesis
$v^i(\vu) \not \equiv 0$ ensures ${\hat v}^i (\vu) \not \equiv 0$.
If $n=3$ and $v^3(\vu) = 0$, then Theorem \ref{abe:teonec1} is not
applicable for transformations of the independent variable $t$.
Indeed, the zero curvature equations for the transformed metric
take the form
\[
{\hat v}^3 (\vu)=0, \quad {\hat n}^3(\vu) =0, \quad {\hat
n}^2(\vu) {\hat v}^1 (\vu)+{\hat n}^1(\vu) {\hat v}^2(\vu) \equiv
0,
\]
instead of (\ref{abe:casodeli}). The condition ${\hat
n}^3(\vu)\equiv 0$ implies $n^3(\vu)\equiv 0$, but we can't
conclude that ${\hat n}^1(\vu)=0={\hat n}^2(\vu)$ and in general
we may get a transformed flat metric with non-canonical
transformations. Indeed, let
\[
u^1_t =2u^1_x,\quad\quad u^2_t =u^2_x,\quad\quad u^3_t =0.
\]
The above system is integrable and possesses a local Hamiltonian
structure associated to the flat metric $g = (du^1)^2 +
(du^2)^2+(du^3)^2$.  Let the reciprocal transformation be $d{\hat
x}=dx$, $d{\hat t} =N(\vu)dx +M(\vu)dt$, with
\[
N(\vu) = \exp(u^1)+\exp(-u^1) +
2\sqrt{2}\cos(\frac{u^2}{2})+2\sin(\frac{u^2}{2})+u^3,
\]
\[
M(\vu)= \exp(u^1)+\exp(-u^1) +
4\sqrt{2}\cos(\frac{u^2}{2})+4\sin(\frac{u^2}{2}).
\]
Then the zero curvature equations for the transformed metric
${\hat g}_{ii} (\vu)$ are identically satisfied and
\[
n^1 (\vu) =\exp(u^1)+\exp(-u^1), \quad n^2(\vu)
=-\frac{\sqrt{2}}{2}\cos(\frac{u^2}{2})-\frac{1}{2}\sin(\frac{u^2}{2}),\quad
n^3 (\vu)=0.
\]
\end{remark}

\medskip

In the following Theorem we settle the necessary conditions for
reciprocal flat metrics in the case of a reciprocal transformation
of both the independent variables.

\begin{theorem}\label{abe:teonec2}
Let $u^i_t =v^i(\vu)u^i_x$, $i=1,\dots,n$, $n\ge
5$, be an integrable strictly hyperbolic DN hydrodynamic type
system as in (\ref{abe:sisdia}), let ${ g}_{ii}(\vu)$ be one of
its metrics with Hamiltonian operator $J^{ij} (\vu)$ as in
(\ref{abe:ham}). Let
\begin{equation}\label{abe:dirtrasf}
d{\hat x} =  B(\vu) d{ x} + A (\vu) d{t},\quad\quad d{\hat t} = N
(\vu) d x + M (\vu) d t,
\end{equation}
be a reciprocal transformation such that the reciprocal metric
${\hat g}_{ii} (\vu)$ defined in (\ref{abe:tramet}) be flat. Then

\smallskip

i) There exist (possibly null) affinors $w^i_{(l)} (\vu)$,
$i=1,\dots, n$, $l=1,2$, such that the Riemannian curvature tensor
of the initial metric $g_{ii} (\vu)$ takes the form
\begin{equation}\label{abe:recriem}
R^{ij}_{ij} (\vu) = w^i_{(1)} (\vu) + w^j_{(1)}  (\vu) + w^i_{(2)}
(\vu) v^{j} (\vu) + w^j_{(2)} (\vu) v^{i} (\vu),\quad\quad i\not
=j;
\end{equation}

ii) the reciprocal transformation (\ref{abe:dirtrasf}) is
canonical with respect to $J^{ij}(\vu)$, the Hamiltonian operator
associated to the initial metric $g_{ii} (\vu)$. In particular,
the auxiliary flows
\[
u^i_{\zeta} = b^i(\vu) u^i_x = J^{ij}(\vu)
\partial_j B(\vu), \quad\quad
u^i_{\tau} = n^i(\vu) u^i_x = J^{ij}(\vu)
\partial_j N(\vu),
\]
associated to such transformations are linear combinations of the
$x$ and $t$ flows.
\end{theorem}

{\sl Proof of the theorem} To verify property i) it is sufficient
to invert the reciprocal transformation (\ref{abe:dirtrasf}) and
to apply Theorem \ref{abe:theo3.2} to the reciprocal flat metric
${\hat g}_{ii} (\vu)$.

We now prove statement ii) in the case of a general reciprocal
transformation (\ref{abe:dirtrasf}) and let the initial metric
$g_{ii} (\vu)$ have Riemann curvature tensor as in
(\ref{abe:recriem}).

Let $n=5$. The zero curvature equations associated to the
reciprocal flat metric ${\hat g}_{ii} (\vu)$ are
\[
{\hat b}^{i} (\vu) + {\hat b}^{j}(\vu) + {\hat n}^i(\vu) {\hat
v}^{j} (\vu) + {\hat n}^j(\vu) {\hat v}^{i} (\vu) =0, \quad\quad
i\not = j.
\]
Using the strict hyperbolicity hypothesis, it is elementary to
show that they may be equivalently expressed as
\[
{\hat b}^i (\vu) = - {\hat n}^{1} (\vu) {\hat v}^i(\vu),
\quad\quad {\hat n^i} (\vu) = {\hat n}^1(\vu), \quad\quad
i=1,\dots, 5.
\]
For $n\ge 5$, it is also easy to prove by induction that the
system of zero curvature equations in the $2n$ variables ${\hat
b}^i(\vu) $, ${\hat n}^i(\vu)$ has rank $2n-1$ and that
\begin{equation}\label{abe:zerocurv1}
{\hat b}^i (\vu) = - {\hat n}^{1} (\vu) {\hat v}^i(\vu),
\quad\quad {\hat n^i} (\vu) = {\hat n}^1(\vu), \quad\quad
i=1,\dots, n.
\end{equation}
Since ${\hat n}^j(\vu)$ are affinors of the transformed metric
${\hat g}_{ii} (\vu)$, using (\ref{abe:commuting}) for the
transformed metric and (\ref{abe:zerocurv1}), we have $\partial_k
{\hat n}^j(\vu)\equiv 0$, $k\not =j$. Using again
(\ref{abe:zerocurv1}), we then conclude that there exists a
(possibly null) constant $\kappa_0$ such that
\begin{equation}\label{abe:zerocurv}
{\hat b}^i (\vu) = - \kappa_0 {\hat v}^i(\vu), \quad\quad {\hat
n^i} (\vu) = \kappa_0, \quad\quad i=1,\dots, n.
\end{equation}
For the inverse reciprocal transformation, we have
\[
\begin{array}{l}
d x = {\hat B}(\vu) d{\hat x} + {\hat A}(\vu) d{\hat t} + {\hat
Q}(\vu) d{\hat \zeta} + {\hat P}(\vu)d{\hat \tau},\\
\\
dt = {\hat N}(\vu) d{\hat x} + {\hat M}(\vu) d{\hat t} + {\hat
S}(\vu) d{\hat \zeta} + {\hat R}(\vu)d{\hat \tau}, \quad \zeta =
{\hat \zeta},\quad \tau ={\hat \tau},
\end{array}
\]
with
\begin{equation}\label{abe:reccons}
\begin{array}{ll}
\displaystyle {\hat B} (\vu) =
\frac{M(\vu)}{B(\vu)M(\vu)-A(\vu)N(\vu)},&\quad\displaystyle {\hat
A} (\vu) = -\frac{A(\vu)}{B(\vu)M(\vu)-A(\vu)N(\vu)},\\
&\\ \displaystyle {\hat N} (\vu) =
-\frac{N(\vu)}{B(\vu)M(\vu)-A(\vu)N(\vu)},&\quad\displaystyle
{\hat M} (\vu) = \frac{B(\vu)}{B(\vu)M(\vu)-A(\vu)N(\vu)},\\ &\\
\displaystyle {\hat Q} (\vu) =
\frac{S(\vu)A(\vu)-Q(\vu)M(\vu)}{B(\vu)M(\vu)-A(\vu)N(\vu)},
&\quad\displaystyle
{\hat S} (\vu) = \frac{Q(\vu)N(\vu)-S(\vu)B(\vu)}{B(\vu)M(\vu)-A(\vu)N(\vu)},\\
&\\ \displaystyle {\hat P} (\vu) =
\frac{R(\vu)A(\vu)-P(\vu)M(\vu)}{B(\vu)M(\vu)-A(\vu)N(\vu)},
&\quad\displaystyle {\hat R} (\vu) =
\frac{P(\vu)N(\vu)-R(\vu)B(\vu)}{B(\vu)M(\vu)-A(\vu)N(\vu)}.
\end{array}
\end{equation}
Since
\[
{\hat v}^i(\vu) =
\frac{B(\vu)v^i(\vu)-A(\vu)}{M(\vu)-N(\vu)v^i(\vu)} =
\frac{\partial_i {\hat A} (\vu)}{\partial_i {\hat B}(\vu)}=
\frac{\partial_i {\hat M} (\vu)}{\partial_i {\hat N}(\vu)}, \quad
\quad i=1,\dots,n,
\]
and the reciprocal affinors satisfy ($i=1,\dots, n$)
\[
{\hat n}^i (\vu) = B (\vu)n^i(\vu) -P (\vu)+
(N(\vu)n^i(\vu)-R(\vu)){\hat v}^i(\vu) = \frac{\partial_i {\hat
P}(\vu) }{\partial_i {\hat B}(\vu)}= \frac{\partial_i {\hat
R}(\vu) }{\partial_i {\hat N}(\vu)},
\]
\[
{\hat b}^i (\vu) = B (\vu)n^i(\vu) -Q (\vu)+
(N(\vu)n^i(\vu)-S(\vu)){\hat v}^i (\vu)= \frac{\partial_i {\hat
Q}(\vu) }{\partial_i {\hat B}(\vu)}= \frac{\partial_i {\hat
S}(\vu) }{\partial_i {\hat N}(\vu)},
\]
we immediately conclude that there exist constants
$\kappa_1,\dots,\kappa_4$ such that
\[
\begin{array}{ll}
{\hat S} (\vu) = -\kappa_0 {\hat M} (\vu)-\kappa_1, &\quad {\hat
Q} (\vu) = -\kappa_0 {\hat A} (\vu)-\kappa_2,\\
&\\
{\hat R} (\vu) = \kappa_0 {\hat N} (\vu)-\kappa_3, &\quad {\hat P}
(\vu) = \kappa_0 {\hat B} (\vu)-\kappa_4.
\end{array}
\]
Inserting (\ref{abe:reccons}), into the above equations, we then
get
\[
\begin{array}{ll} Q(\vu) = \kappa_2 B(\vu)+\kappa_1 A(\vu)  &\quad S(\vu) =
\kappa_2 N(\vu) +\kappa_1 M(\vu)+\kappa_0,
\\
&\\
P(\vu) = \kappa_4 B(\vu) +\kappa_3 A(\vu) -\kappa_0, &\quad R(\vu)
= \kappa_4 N(\vu) +\kappa_3 M(\vu),
\end{array}
\]
from which the assertion follows. $\quad \square$

\medskip

\begin{remark}
If $n=4$, the system of the six zero curvature equations for the
transformed metric ${\hat g}_{ii} (\vu)$ has maximal rank 6 in the
unknowns ${\hat b}^i$, ${\hat n}^i$, and it is possible to
express, say ${\hat b}^i (\vu)$, ${\hat n}^i (\vu)$, $i=2,3,4$ in
function of ${\hat b}^1(\vu)$ and ${\hat n}^1(\vu)$. Moreover the
condition ${\hat n}^i(\vu)={\hat n}^1(\vu)$, $i=2,3,4$ is
satisfied if and only if ${\hat b}^1 (\vu) = -{\hat v}^1 (\vu)
{\hat n}^1 (\vu)$.

The above observation implies that, for $n=4$, there exist
non-canonical transformations which preserve the flatness of the
metric when ${\hat n}^i(\vu)\not ={\hat n}^1(\vu)$ for $i\in
\{2,3,4\}$.
\end{remark}

\section{Classification of the reciprocal transformations which
preserve the flatness of the metric or transform constant
curvature metrics to flat metrics}

Theorems \ref{abe:teonec1} and \ref{abe:teonec2} state that only
the reciprocal transformations which are canonical with respect to
the initial Hamiltonian structure may transform the initial metric
to a flat one, respectively for $n\ge 3$ (reciprocal
transformations of a single independent variable) or $n\ge 5$
(reciprocal transformations of both the independent variables). In
view of the above, in this section we restrict ourselves to the
case in which the initial metric $g_{ii} (\vu)$ is either flat
($w^i_{(1)}\equiv 0 \equiv w^i_{(2)}$, $i=1,\dots,n$, in
(\ref{abe:recriem})) or of constant curvature $2c$
($w^i_{(1)}\equiv c$, $w^i_{(2)}\equiv 0$, $i=1,\dots,n$, in
(\ref{abe:recriem})). Then, in Theorem \ref{abe:teosuff},  we
completely characterize which reciprocal transformations map
$g_{ii} (\vu)$ to flat metric ${\hat g}_{ii} (\vu)$.

Finally, the case in which both the initial and the transformed
Hamiltonian structure are local has also a nice geometric
interpretation in view of the results by Ferapontov
\cite{abe:F95-1}, which we present in Theorem \ref{abe:cor}.

\begin{theorem}\label{abe:teosuff}
Let $n\ge 5$ and let $u^i_t =v^i(\vu) u^i_x=J^{ij} (\vu)
\partial_j H(\vu)$, $i=1,\dots, n$, be a DN integrable strictly
hyperbolic hydrodynamic type system, with $J^{ij} (\vu)$ the
Hamiltonian operator associated to the metric $g_{ii} (\vu)$ and
$H(\vu)$ the Hamiltonian density. Let $d{\hat x} = B(\vu) dx +
A(\vu)dt$, $d{\hat t} = N(\vu) dx + M(\vu)dt$ be a reciprocal
transformation with $A(\vu),B(\vu),M(\vu)$ and $N(\vu)$ not all
constant functions.

\smallskip

A) Let the metric $g_{ii} (\vu)$ be flat. Then the reciprocal
metric ${\hat g}_{ii} (\vu)$ defined in (\ref{abe:tramet}) is flat
if and only if one of the following alternatives hold:

A.i) there exist constants $\kappa_1\not =0,\kappa_2,\kappa_3$
such that
\[
M(\vu)=\kappa_1, \quad\quad N(\vu) = \kappa_2,\quad\quad \left(
\nabla B \right)^2 (\vu) = \kappa_3\left(\kappa_1 B(\vu)-\kappa_2
A(\vu)\right);
\]

A.ii) there exist constants $\kappa_1\not =0,\kappa_2,\kappa_3$
such that
\[
B(\vu)=\kappa_1, \quad\quad A(\vu) = \kappa_2,\quad\quad \left(
\nabla N \right)^2 (\vu) = \kappa_3\left( \kappa_1 M(\vu)-\kappa_2
N(\vu)\right);
\]

A.iii) there exist constants $\kappa_1,\kappa_2,\kappa_3,\kappa_4$
such that
\[
\left( \nabla B \right)^2 (\vu) = 2\kappa_1 A(\vu) +2 \kappa_2
B(\vu), \quad\quad \left( \nabla N \right)^2 (\vu) = 2 \kappa_3
M(\vu)+2\kappa_4 N(\vu),\]
\[
<\nabla B(\vu), \nabla N(\vu) > = \kappa_1 M(\vu) + \kappa_2
N(\vu) +\kappa_3 A(\vu) +\kappa_4 B(\vu).
\]

\smallskip

B) Let the metric $g_{ii} (\vu)$ be of constant curvature $2c$.
Then the reciprocal metric ${\hat g}_{ii} (\vu)$ defined in
(\ref{abe:tramet}) is flat if and only if one of the following
alternatives hold:

B.i) there exist constants $\kappa_1\not =0,\kappa_3,$ such that
\[
M(\vu)=\kappa_1, \quad\quad N(\vu) \equiv 0,\quad\quad \left(
\nabla B \right)^2 (\vu) + 2cB^2(\vu)= 2\kappa_3 B(\vu);
\]

B.ii) there exist constants $\kappa_1,\kappa_2,\kappa_3,\kappa_4$
such that
\[
\left( \nabla B \right)^2 (\vu) + 2cB^2(\vu) = 2\kappa_1 A(\vu) +2
\kappa_2 B(\vu), \quad\quad \left( \nabla N \right)^2 (\vu)
+2cN^2(\vu) = 2 \kappa_3 M(\vu)+2\kappa_4 N(\vu),\]
\[
<\nabla B(\vu), \nabla N(\vu) > +2cB(\vu)N(\vu)= \kappa_1 N(\vu) +
\kappa_2 M(\vu) +\kappa_3 A(\vu) +\kappa_4 B(\vu).
\]
\end{theorem}

\medskip

\begin{remark}
Case A.i) (resp. A.ii) ) includes the reciprocal transformations
of the single variable $x$ (resp. the single variable $t$) when
$\kappa_1 =1, \kappa_2 =0$.

Case B.i) corresponds to reciprocal transformations of the single
variable $x$ (notice that only $N(\vu)\equiv 0$ is admissible if
$c\not =0$). Finally, it is not possible to transform a constant
curvature metric to a flat one by a transformation of the single
variable $t$.
\end{remark}

\medskip

{\sl Proof:} Let $g_{ii} (\vu)$ be either a flat ($c=0$) or a
constant curvature metric ($c\not =0$).

We prove first A.i) and B.i). Let $\kappa_1\not =0,\kappa_2$ be
constants such that
\[
M(\vu) \equiv \kappa_1,\quad\quad N(\vu) \equiv \kappa_2.
\]
Then, the only possibly non-zero elements of the reciprocal
Riemannian curvature tensor take the form,
\[
{\hat R}^{ik}_{ik} (\vu) = 2c\frac{H_i(\vu)H_k(\vu)}{{\hat
H}_i(\vu){\hat H}_k(\vu)}-\left(\nabla B \right)^2 (\vu)
+\frac{H_i(\vu)}{{\hat H}_i (\vu)} \nabla^i \nabla_i
B(\vu)+\frac{H_k(\vu)}{{\hat H}_k (\vu)} \nabla^k \nabla_k B(\vu),
\]
where
\[
{\hat H}_i(\vu) = \frac{\kappa_1-\kappa_2 v^i(\vu)}{B(\vu)
\kappa_1 -A(\vu)\kappa_2}H_i (\vu), \quad\quad {\hat v}^i (\vu)
=\frac{B(\vu)v^i(\vu)-A(\vu)}{\kappa_1 -\kappa_2
v^i(\vu)},\quad\quad i=1,\dots,n.
\]
From the necessary condition found in Theorems \ref{abe:teonec1}
and \ref{abe:teonec2},
\begin{equation}\label{abe:1}
b^i(\vu) \equiv \nabla^i\nabla_i B(\vu) +2cB(\vu)
=\kappa_3+\kappa_4 v^i(\vu), \quad\quad i=1,\dots,n,
\end{equation}
we infer
\begin{equation}\label{abe:2}
\left(\nabla B\right)^2 (\vu) +2cB^2 (\vu)= 2\kappa_3 B(\vu)
+2\kappa_4 A(\vu) +\kappa_5.
\end{equation}
If we insert (\ref{abe:1}) and (\ref{abe:2}) inside the expression
of the transformed Riemannian curvature tensor, we immediately get
\[
{\hat R}^{ik}_{ik} (\vu) = -\kappa_5 +(\kappa_1\kappa_4
+\kappa_3\kappa_2) \big({\hat v}^i(\vu) +{\hat v}^k (\vu)\big)
+2c\kappa_2^2 {\hat v}^i(\vu)  {\hat v}^k(\vu).
\]
Then the condition ${\hat R}^{ik}_{ik} (\vu)\equiv 0$, is
equivalent to either
\[
c=\kappa_5=\kappa_1\kappa_4 +\kappa_3\kappa_2=0,
\]
or to
\[
c\not =0, \quad \mbox{and} \quad \kappa_5=\kappa_2=\kappa_4=0,
\]
from which cases  A.i) and B.i) immediately follow.
\smallskip

We now prove A.ii). Let $\kappa_1\not =0,\kappa_2$ be constants
such that $B(\vu) \equiv \kappa_1$,$A(\vu) \equiv \kappa_2$ and
let the initial metric $g_{ii} (\vu)$ be flat. Then, the only
possibly non-zero elements of the reciprocal Riemannian curvature
tensor take the form,
\[
{\hat R}^{ik}_{ik} (\vu) =\frac{H^i(\vu)}{{\hat H}^i (\vu)}
\nabla^i\nabla_i N(\vu) {\hat v}^k (\vu) + \frac{H^k(\vu)}{{\hat
H}^k (\vu)} \nabla^k\nabla_k N(\vu) {\hat v}^i (\vu) -{\hat
v}^i(\vu) {\hat v}^k (\vu) \left( \nabla N\right)^2 (\vu),\quad
i\not =k.
\]
Inserting into the above equation ($i=1,\dots,n$)
\[
{\hat H}_i (\vu) = \frac{M(\vu)-N(\vu) v^i (\vu)}{\kappa_1 M(\vu)
-\kappa_2 (\vu) N(\vu)} H_i (\vu), \quad \quad {\hat v}^i(\vu) =
\frac{\kappa_1 v^i(\vu)-\kappa_2}{M(\vu)-N(\vu) v^i(\vu)},
\]
we get
\[
{\hat R}^{ik}_{ik} (\vu) = {\hat v}^i(\vu) {\hat v}^k(\vu) \left(
\frac{(\kappa_1 M(\vu)-\kappa_2N(\vu)n^i (\vu)}{\kappa_1 v^i
(\vu)-\kappa_2}+ \frac{(\kappa_1 M(\vu)-\kappa_2N(\vu)n^k
(\vu)}{\kappa_1 v^k (\vu)-\kappa_2}- \big( \nabla N\big)^2\right).
\]
Since ${\hat v}^i(\vu)\not \equiv 0$, the conditions ${\hat
R}^{ik} (\vu)\equiv 0$, ($i\not=k$) are equivalent to either
$(\nabla N)^2 (\vu)\equiv 0$ or
\[
\frac{\displaystyle \partial_i \left(\nabla N\right)^2
(\vu)}{\left(\nabla N\right)^2 (\vu)} = \frac{\displaystyle
\partial_i \left(\kappa_1 M(\vu) -\kappa_2 N(\vu)\right)}{
\left(\kappa_1 M(\vu) -\kappa_2 N(\vu)\right)},\quad\quad
i=1,\dots, n,
\]
from which case {\it ii)} immediately follows. In particular,
under the same hypotheses, we also have
\[
n^i(\vu) = \kappa_3 (\kappa_1 v^i(\vu)+\kappa_2), \quad\quad {\hat
n}^i(\vu) = -\frac{1}{2} \left( \nabla N\right)^2 (\vu) {\hat
v}^i(\vu) +n^i(\vu) \frac{\kappa_1
M(\vu)-\kappa_2N(\vu)}{M(\vu-N(\vu)v^i(\vu)}\equiv 0.
\]
If $\kappa_1\not =0,\kappa_2$ are constants such that $B(\vu)
\equiv \kappa_1$,$A(\vu) \equiv \kappa_2$ and the initial metric
$g_{ii}(\vu)$ is of constant curvature $c\not=0$, then it is easy
to show that the transformed metric ${\hat g}_{ii}$ cannot be
flat.

\smallskip
To prove {\it A.iii)} and {\it B.ii)}, we use the closed form
\begin{equation}\label{abe:clofor1}
\left\{ \begin{array}{l} \displaystyle d{\hat x} = B (\vu)dx +
A(\vu)dt+ P(\vu)d\tau + Q(\vu) d\zeta,
\\\displaystyle d{\hat t} = N(\vu)dx + M(\vu)dt+R(\vu)d\tau+
S(\vu)d\zeta,\\
\displaystyle d{\hat \tau} = d\tau, \quad d{\hat \zeta} = d\zeta,
\end{array}\right.\end{equation}
associated to the auxiliary flows
\begin{equation}\label{abe:bn}
u^i_{\tau} = n^{i} (\vu) u^i_x = \left({ \nabla}^i{ \nabla}_i
N(\vu) +2cN(\vu)\right) u^i_x,\quad u^i_{\zeta} = b^{i} (\vu)
u^i_x= \left( { \nabla}^i{ \nabla}_i B (\vu)
+2cB(\vu)\right)u^i_x.
\end{equation}
In view of the results of the previous section, the auxiliary
flows (\ref{abe:bn}) are necessarily linear combinations of the
$x$ and $t$ flows. We impose that the conservation laws in the
reciprocal transformation satisfy the necessary conditions settled
in Theorem \ref{abe:teonec2}. Then there exist constants
$\kappa_j$, $j=1,\dots, 8$ such that
\[
\begin{array}{ll}
b^i(\vu) =\kappa_1 v^i(\vu) + \kappa_2,&\quad \left(\nabla
B\right)^2 (\vu) +2cB(\vu)= 2\kappa_1 A(\vu) +2\kappa_2 B(\vu)
+2\kappa_6,\\
n^i(\vu) =\kappa_3 v^i (\vu) +\kappa_4, &\quad \left(\nabla
N\right)^2 (\vu) +2cN(\vu)= 2\kappa_3 M(\vu) +2\kappa_4 N(\vu)
+2\kappa_5,\\ P(\vu) = \kappa_3 A(\vu) +\kappa_4 B(\vu)
+\kappa_7,&\quad S(\vu) =\kappa_1 M(\vu) +\kappa_2 N(\vu)
+\kappa_8,
\end{array}
\]
\[<\nabla B,\nabla N > +2c BN\equiv
P+S = \kappa_1 M(\vu) +\kappa_2 N(\vu) + \kappa_3 A(\vu) +\kappa_4
B(\vu) +\kappa_7 +\kappa_8.
\]
If we insert the above expressions into the right hand side of
(\ref{abe:travel}) we get
\[
{\hat n}^i (\vu) = -\kappa_7 -\kappa_5 {\hat v}^i(\vu),\quad\quad
{\hat b}^i(\vu) = -\kappa_6-\kappa_8 {\hat v}^i(\vu).
\]
Finally, the elements of the Riemannian curvature tensor are
\[
\begin{array}{ll} {\hat R}^{ik}_{ik} (\vu) &= {\hat n}^i (\vu){\hat v}^i(\vu)+
{\hat n}^k (\vu) {\hat v}^k(\vu) +{\hat b}^i(\vu) +{\hat b}^k(\vu)
\\
&= -2\kappa_6 -(\kappa_7 +\kappa_8) ({\hat v}^i(\vu) +{\hat
v}^k(\vu) ) -2\kappa_5 {\hat v}^i (\vu) {\hat v}^k(\vu),\quad
i\not =k,
\end{array}
\]
so that ${\hat R}^{ik}_{ik} (\vu) \equiv 0$ if and only if
$\;\kappa_5=\kappa_6=\kappa_7+\kappa_8=0$, and the assertions {\it
A.iii)} and {\it B.ii)} easily follow. $\quad\quad \square$

\medskip

\begin{example}
If $B(\vu)$ and $N(\vu)$ are non trivial independent Casimirs of
the flat metric $g_{ii} (\vu)$ and $(\nabla B (\vu))^2\not =0$,
then there exist a constant $\alpha$ and $A(\vu)$ such that, under
the reciprocal transformation $d{\hat x} = (\alpha B(\vu)+N(\vu))
dx + A(\vu) dt$, the reciprocal metric ${\hat g}_{ii} (\vu) =
g_{ii} (\vu)/(\alpha B(\vu)+N(\vu))^2 $ is flat.
\end{example}

\begin{example}
If $B(\vu)$ is a density of momentum for the flat metric $g^{ii}
(\vu)$ and $(\nabla B(\vu))^2-2B(\vu)=2\alpha$, then under the
reciprocal transformation $d{\hat x} = (B(\vu)+\alpha) dx + A(\vu)
dt$, the reciprocal metric ${\hat g}_{ii} (\vu) = g_{ii}
(\vu)/(B(\vu)+\alpha)^2 $ is flat.
\end{example}

\begin{example}
If $B(\vu)$ and $N(\vu)$ are non trivial independent Casimirs of
the flat metric $g_{ii} (\vu)$ and $(\nabla N (\vu))^2\not =0$,
then there exist a constant $\alpha$ and $M(\vu)$ such that, under
the reciprocal transformation $d{\hat t} = (\alpha N(\vu)+B(\vu))
dx + M(\vu) dt$, the reciprocal metric ${\hat g}_{ii} (\vu)$ is
flat.
\end{example}

\begin{example}
If $N(\vu)$ is a density of Hamiltonian for the flat metric
$g_{ii} (\vu)$ and $(\nabla N(\vu))^2-2M(\vu)=2\alpha$, then under
the reciprocal transformation $d{\hat t} = N(\vu) dx +
\Big(M(\vu)+\alpha\big) dt$, the reciprocal metric ${\hat g}_{ii}
(\vu) $ is flat.
\end{example}

\begin{example}
Let $N(\vu)$ be a density of momentum and let $B(\vu)$ be a
density of Hamiltonian for the flat metric $g_{ii} (\vu)$. Then
under the reciprocal transformation
\[
d{\hat x} = B(\vu)dx + \frac{1}{2} (\nabla B)^2 (\vu) dt,
\quad\quad d{\hat t} = N(\vu) dx +M(\vu) dt,
\]
such that $(\nabla N)^2 (\vu)= 2N(\vu)$, $< \nabla N(\vu), \nabla
B(\vu)
> = N(\vu) + B(\vu)$, the reciprocal metric ${\hat g}_{ii}(\vu)$ is flat.
\end{example}

\begin{example}
Let $N(\vu)=M(\vu)=1$ and let $B(\vu)$ be a density of Hamiltonian
for the flat metric $g_{ii} (\vu)$. Then under the reciprocal
transformation
\[
d{\hat x} = B(\vu)dx + \frac{1}{2} (\nabla B)^2 (\vu) dt,
\quad\quad d{\hat t} = dx + dt,
\]
the reciprocal metric ${\hat g}_{ii}(\vu)$ is flat.
\end{example}

\begin{example}
Let $N(\vu)$ be a density of momentum and let $B(\vu)$ be a
density of Hamiltonian for the  metric $g_{ii} (\vu)$ with
constant curvature $2c$. Then under the reciprocal transformation
\[
d{\hat x} = B(\vu)dx + \left(\frac{1}{2} (\nabla B)^2 (\vu) +c
B^2(\vu)\right) dt, \quad\quad d{\hat t} = N(\vu) dx +M(\vu) dt,
\]
such that $(\nabla N)^2 (\vu) + 2c N^2 (\vu)- 2N(\vu)\equiv 0$, $<
\nabla N(\vu), \nabla B(\vu)
> +2c N(\vu)B(\vu)- N(\vu) - B(\vu)\equiv 0$, the reciprocal metric
${\hat g}_{ii}(\vu)$ is flat.
\end{example}

\begin{example}
Let $N(\vu)$ be a density of Hamiltonian and let $B(\vu)$ be a
Casimir for the  metric $g_{ii} (\vu)$ with constant curvature
$2c$. Then under the reciprocal transformation
\[
d{\hat x} = B(\vu)dx + A(\vu) dt, \quad\quad d{\hat t} = N(\vu) dx
+\left(\frac{1}{2} (\nabla N)^2 (\vu) +c N^2(\vu)\right) dt,
\]
such that $(\nabla B)^2 (\vu) + 2c B^2 (\vu)\equiv 0$, $< \nabla
N(\vu), \nabla B(\vu)
> +2c N(\vu)B(\vu)- B(\vu)\equiv 0$, the reciprocal metric
${\hat g}_{ii}(\vu)$ is flat.
\end{example}

\subsection{Reciprocal transformations which preserve the flatness property
of the metric and Lie--equivalent systems}

We end the paper giving the geometrical interpretation of Theorem
\ref{abe:teosuff} in the case in which both the initial and the
transformed metrics are flat. Indeed, local Hamiltonian systems
connected by canonical reciprocal transformations have nice
geometrical properties as first observed by Ferapontov
\cite{abe:F95-1}. Using the theorems proven by Ferapontov in
\cite{abe:F95-1} and Theorem \ref{abe:teosuff}, in Theorem
\ref{abe:cor} we show that the local Hamiltonian structures of two
DN Hamiltonian systems in Riemann invariants are connected by a
canonical reciprocal transformation if and only if the associated
hypersurfaces are Lie equivalent.

A DN hydrodynamic type system as in (\ref{abe:DN}) in flat
coordinates takes the form
\begin{equation}\label{abe:hamflat}
u^i_t = v^{i}_j (\vu) u^i_x = \epsilon^i \delta^{ij}\frac{d}{dx}
\left( \frac{\delta H}{\delta u^j}\right),
\end{equation}
with $\epsilon^i=\pm 1$ and the Hamiltonian $H= \int h(\vu) dx$.
To each system as in (\ref{abe:hamflat}), there corresponds a
hypersurface $M^n$ in a pseudoeuclidean space $E^{n+1}$ in such a
way that equations (\ref{abe:hamflat}) may be transformed into the
form
\begin{equation}\label{abe:fernr}
\no_t = \ru_x, \end{equation} where $\no$ and $\ru$ are
respectively the unit normal and the radius vector of $M^n$ (see
\cite{abe:F95-1}). Let $u^1,\dots, u^n$ be any system of
curvilinear coordinates on $M^n$. Since the tangent bundle $TM^n$
is spanned by $\frac{\partial \ru}{\partial u^i}$, $i=1,\dots,n$
and $\frac{\partial \no}{\partial u^i}\in TM^n$, $i=1,\dots,n$, it
is possible to introduce the so-called Weingarten (or shape)
operator $w^i_j(\vu)$, by the formulas
\[
\frac{\partial \no}{\partial u^j} = w^i_j(\vu) \frac{\partial
\ru}{\partial u^j},
\]
and (\ref{abe:fernr}) may be rewritten in the form
(\ref{abe:hamflat}), with $v^i_j = (w^i_j)^{-1}$. Then the
eigenvalues of the velocities $v^i_j(\vu)$ are the radii of the
principal curvatures of $M^n$ and the corresponding
eigenfoliations are the curvature surfaces of $M^n$ (see
\cite{abe:F95-1}). In particular, the hypersurface $M^n$ is called
Dupin if its principal curvatures are constant along the
corresponding curvature hypersurfaces and such hypersurfaces
correspond to weakly--nonlinear hydrodynamic type systems ({\it
i.e.} each eigenvalue of the matrix $v^{ij} (\vu)$ in
(\ref{abe:hamflat}) is constant along the corresponding
eigenfolation) as proven in \cite{abe:F95-1}.

Following \cite{abe:F95-1}, let us call the hypersurfaces
associated to two DN systems as in (\ref{abe:hamflat})
Lie--equivalent if they are connected by a Lie sphere
transformation (see \cite{abe:Lie}, \cite{abe:Cec}).

The $n+2$--canonical integrals (the $n$ Casimirs, the momentum and
the Hamiltonian) take the following form in the flat coordinates
$u^1,\dots, u^n$ (see \cite{abe:F95-1}),
\[
\begin{array}{c}
\displaystyle H = hdx + \frac{1}{2} \Big(\sum_{m=1}^n \epsilon^m
(\partial_m h)^2 +1\Big)
dt,\\
\displaystyle P = \frac{1}{2}\Big(\sum_{m=1}^n \epsilon^m u_m^2
+1\Big) dx - \Big(h-\sum_{m=1}^n u_m\partial_m h\Big) dt,\\
\displaystyle U^i = u^i dx +\epsilon^i \partial_i h dt, \quad
i=1,\dots, n.
\end{array}
\]
Then the following Theorem settles the following important
relation between equivalent hypersurfaces and reciprocal
transformations.

\begin{theorem}\label{abe:teofer} \cite{abe:F95-1}

A) Suppose that the associated hypersurfaces of two DN systems as
in (\ref{abe:hamflat}) are Lie--equivalent. Then the local
Hamiltonian structures of the systems themselves are connected by
a reciprocal transformation.

B) Suppose that the local Hamiltonian structures of two DN systems
are connected by the  canonical reciprocal transformation
\[
d{\hat x} = \alpha H + \beta P + \sum_{m=1}^n \gamma_i U^i +
\eta_1 dx + \eta_2 dt,\quad\quad d{\hat t} = {\tilde \alpha} H +
\beta P + \sum_{m=1}^n {\tilde \gamma_i} U^i + {\tilde \eta_1} dx
+ {\tilde \eta_2} dt, \] with
$\alpha,\beta,\gamma_m,\eta_j,{\tilde \alpha},{\tilde \beta},{
\tilde \gamma}_m,{\tilde \eta}_j$, ($m=1,\dots, n$, $j=1,2$)
constants such that
\begin{equation}\label{abe:compfer}
\begin{array}{c}
\displaystyle (\alpha + \eta_1)^2 + (\beta + \eta_2)^2
-\sum_{m=1}^n \epsilon^m \gamma_m^2 - \eta_1^2-\eta_2^2 =0,\\
\displaystyle ({\tilde \alpha} + {\tilde\eta_1})^2 + ({\tilde
\beta} + {\tilde \eta}_2)^2 -\sum_{m=1}^n \epsilon^m {\tilde
\gamma}_m^2 - {\tilde \eta}_1^2-
{\tilde \eta}_2^2 =0,\\
\displaystyle ({\tilde \alpha} + {\tilde\eta_1})(\alpha + \eta_1)
+ ({\tilde \beta} + {\tilde \eta}_2)(\beta + \eta_2) -\sum_{m=1}^n
\epsilon^m {\tilde \gamma}_m\gamma_m - {\tilde \eta}_1\eta_1-
{\tilde \eta}_2\eta_2 =0.
\end{array}
\end{equation}
Then the hypersurfaces associated to the two DN systems are
Lie--equivalent.
\end{theorem}

We recall that any $n\times n$ DN type system as in
(\ref{abe:hamflat}) admits the $n+2$--canonical integrals, so
that Theorem \ref{abe:teofer} applies also to the case in which
Riemann invariants do not exist.

If we restrict ourselves to the case of DN systems which possess
Riemann invariants, then the compatibility conditions
(\ref{abe:compfer}) in the flat coordinates have their
correspondence in the conditions A.i)-A.iii) expressed in the
Riemann invariants in Theorem \ref{abe:teosuff}.

Moreover, the same theorem gives the complete characterization of
the reciprocal transformations which preserve local Hamiltonian
structure when Riemann invariants exist, so that the following
stronger geometrical characterization holds in the present case.

\begin{theorem}\label{abe:cor}
Let $n\ge 5$. The hypersurfaces associated to two diagonalizable
strictly hyperbolic DN systems are connected by a Lie sphere
transformation if and only if the corresponding local Hamiltonian
structures of the two DN systems are connected by canonical
reciprocal transformation satisfying Theorem \ref{abe:teosuff}.
\end{theorem}

Finally, we like to point out that there is no geometrical
interpretation of the reciprocal transformations when the locality
of the Hamiltonian structure is not preserved by the
transformation and both the initial and the transformed systems
are of DN type. The most interesting example in this class are the
genus $g$ modulated Camassa-Holm equations already mentioned in
the introduction: such system possesses two compatible flat
metrics which are mapped to two non--flat metrics associated to
the $g$ modulated equations of the first negative Korteweg--de
Vries flow by a reciprocal transformation as proven in
\cite{abe:AG07}. Then, from Theorem \ref{abe:teofer}, it follows
that the hypersurfaces associated to the two systems are not
Lie--equivalent.

\section*{Acknowledgements}
This research has been partially supported by ESF Programme
MISGAM, by RTN ENIGMA and by PRIN2006 ''Metodi geometrici nella
teoria delle onde non lineari ed applicazioni''. I warmly thank
F.V. Ferapontov, M. Pavlov and Y. Zhang for their interest in the
present research. I am also grateful to to the referees for many
useful remarks and especially to the second referee for many
valuable observations which helped me in improving the manuscript.

\end{document}